\documentclass[%
 aip,twocolumn,
 sd,%
amsmath,amssymb,
preprint%
]{revtex4-1}

\usepackage{graphicx}
\usepackage{dcolumn}
\usepackage{bm}
\usepackage{color}
\usepackage{caption}

\newcommand{\lam}{\lambda}

\newcommand{\pa}{\partial}

\begin{document}


\title[]{Fast-Forward Assisted STIRAP}
\author{Shumpei Masuda}
 \altaffiliation[]{Department of Physics, Tohoku University, 
Sendai 980, Japan}
 \email{masuda@uchicago.edu}
\author{Stuart A.  Rice}%
 \email{s-rice@uchicago.edu}
\affiliation{ James Franck Institute, 
The University of Chicago, Chicago, IL 60637 
}%

\date{\today}

\begin{abstract}
We consider combined stimulated Raman adiabatic passage (STIRAP) and 
fast-forward field (FFF) control of selective vibrational population 
transfer in a polyatomic molecule.  The motivation for using this 
combination control scheme is twofold: (i) to overcome transfer 
inefficiency that occurs when the STIRAP fields and pulse durations must 
be restricted to avoid excitation of population transfers that compete with 
the targeted transfer and (ii) to overcome transfer inefficiency resulting 
from embedding of the actively driven subset of states in a large manifold 
of states.  We show that, in a subset of states that is coupled to background 
states, a combination of STIRAP and FFFs that do not individually 
generate processes that are competitive with the desired population transfer 
can generate greater population transfer efficiency than can ordinary STIRAP 
with similar field strength and/or pulse duration.  
The vehicle for our considerations is enhancing the yield of HNC 
in the driven ground state-to-ground state nonrotating HCN $\rightarrow$ HNC 
isomerization reaction and selective population of one of a pair of near degenerate states in nonrotating SCCl$_2$.

\end{abstract}

\maketitle

\section{Introduction}
It is now well established that it is possible to actively control the 
quantum dynamics of a system by manipulating the frequency, phase and 
temporal character of an applied optical field \cite{Rice,Shapiro}.  
The underlying mechanisms of all the proposed and experimentally 
demonstrated active control methods rely on coherence and interference 
effects embedded in the quantum dynamics.  Although the various control protocols provide prescriptions for the calculation of the control field, in general, the manifold of 
states of the driven system is too complicated to permit exact calculation 
of that field. That difficulty has led to the consideration of control of the 
quantum dynamics with a simplified Hamiltonian, e.g. within a subset of states without regard for the influence 
of the remaining background states.  One example of this class 
of control methods is the use of stimulated Raman adiabatic passage 
(STIRAP) \cite{Gaubatz,Coulston,Halfmann,Bergmann,Vitanov} to transfer 
population within a three state subset of a larger manifold of states.  
Various extended STIRAP
methods, involving more than three states, also have been proposed \cite{Kobrak2,Kobrak3,Kobrak1,Kurkal1,Kurkal2,Torosov}.
This simplification is not always acceptable: when the transition dipole 
moments between a selected subset of states and the other (background) 
states of the manifold are not negligible it is necessary to account for 
the influence of transitions involving the background states on the 
efficiency of the population transfer.  Furthermore, STIRAP relies on 
adiabatic driving in which the populations in the instantaneous 
eigenstates of the Hamiltonian are constant.  Because an adiabatic 
process must be carried out very slowly, at a rate much smaller than 
the frequencies of transitions between states of the Hamiltonian, the 
field strength and/or pulse duration imposed must be restricted to avert 
unwanted processes. 
Recognition of this 
restriction has led to the development of control protocols which we call 
assisted adiabatic transformations; these transformations typically use 
an auxiliary field to produce, with overall weaker driving fields and/or 
in a shorter time, and without excitation of competing processes, 
the desired target state population.

In an early study of a version of assisted adiabatic population transfer,
Kurkal and Rice\cite{Kurkal1} used the extended STIRAP process devised by Kobrak and Rice\cite{Kobrak1} to study vibrational
energy transfer between an initial state and two nearly degenerate states in nonrotating
SCCl$_2$. The extended STIRAP process, which is designed to control the ratio of the populations
transferred to the target states, uses three pulsed fields: a pump field, a Stokes field, and a field that couples the target state to a so-called branch
state. The ratio of populations of the target states that can be achieved depends on, and is limited
by, the ratio of the dipole transition moments between the branch state and the target states, and
is discretely controllable by suitable choice of the branch state from the manifold of states. Because
the extended STIRAP process exploits adiabatic population transfer, the field strengths and pulse
durations used must satisfy the same constraints as for a simple adiabatic population transfer.

Other assisted adiabatic transformation control methods include 
the counter-diabatic protocol \cite{Demirplak1,Demirplak2,Demirplak3,Chen}, the invariant-based 
inverse engineering protocol \cite{Muga} and the fast-forward protocol 
\cite{Masuda1,Masuda2,Masuda3,Masuda4,Masuda-Rice1}.

We have shown elsewhere \cite{Masuda-Rice2} that, 
in a subset of states that is coupled to background states, a combination 
of STIRAP fields and a counter-diabatic field (CDF) can generate greater 
population transfer efficiency than can ordinary STIRAP with necessarily 
restricted field strength and/or pulse duration.  And it has been shown 
that the exact CDF for an isolated three level system is a useful 
approximation to the CDF for three and five state sub-manifolds embedded 
in a large manifold of states.

In this paper we complement our previous study with an examination of the use of combined phase-controlled STIRAP 
and fast-forward fields (FFFs) to control selective vibrational population 
transfer in a polyatomic molecule under conditions that require restriction 
of the STIRAP field strength and/or pulse duration. 
Again using selective vibrational energy transfer to drive the rotationless HCN $\rightarrow$ HNC 
isomerization reaction and state-to-state
vibrational energy transfer in an isolated nonrotating
SCCl$_2$ molecule as vehicles for our study it is shown that the phase-controlled STIRAP + 
FFF that affects complete transfer of population in an isolated three-level 
system is a useful approximation to the control field that affects 
efficient
transfer of population for a three-state system embedded in background states.
 The FFF suppresses the influence of background states strongly coupled to 
the STIRAP pumped subset of states.

\section{Fast-Forward Assisted STIRAP}
The fast-forward protocol is constructed to control the rate of evolution 
of particles between selected initial and target states in a continuous 
system.  It can be regarded as defining a trajectory in the state space connecting the initial and final 
states for which the control field that accelerates the initial-to-final 
state transition is realizable.  The time-dependent intermediate states 
acquire, relative to the 
states along the original trajectory of the 
initial-to-final state transition without the FFF acceleration \cite{Masuda1} or the
adiabatic transition \cite{Masuda2}, an additional time-dependent 
phase.  The fast-forward protocol has been extended to treat spatially 
discrete systems, e.g. accelerated manipulation of a Bose-Einstein 
condensate (BEC) in an optical lattice by Masuda and Rice \cite{Masuda-Rice1}, 
and spin systems \cite{Takahashi}, and a variant of this method can be used to 
accelerate selective population transfer between states in a discrete 
spectrum of states of a molecule.

\subsection{Fast-Forward Protocol for Discrete Systems}
We consider a manifold of discrete states $\{|i\rangle\}$ and time 
dependent transition (hopping) rates $\omega_{l,m}$ between states 
$|l\rangle, |m\rangle
\in \{|i\rangle\}$.  Note that these transition rates depend on the 
applied field and are the analogues of the Stokes and Raman 
frequencies
in a STIRAP process; they are not the conventional transition probabilities.  
The derivation of the fast-forward driving fields proceeds in the same 
manner as described in Ref. 23.  
The equation of motion of the system wave function takes the form 
\begin{eqnarray}
i\frac{d\Psi(m,t)}{dt} &=& \sum_l\omega_{m,l}(R(t))\Psi(l,t)\nonumber\\
&&+\frac{V_0(m,R(t))}{\hbar} \Psi(m,t),
\label{SE1}
\end{eqnarray}
where $\Psi(m,t)$ is the coefficient of $|m\rangle$ and $R$ 
is a time-dependent parameter characterizing 
the temporal dependence of $\omega_{m,l}$.  
Eq. (\ref{SE1}) describes the population transfer among molecular states with 
$\omega_{m,l}$ corresponding to the Rabi frequency of the laser 
field coupling the states $|l\rangle$ and $|m\rangle$ and $V_0(m)$ 
the energy of the field-free state $|m\rangle$.  Hereafter we refer 
to $V_0(m)$ as a potential.  Now let $\phi_n(m,R)$ and $E_n(R)$ be the wave 
function (coefficient of $|m\rangle$) and energy of the $n$th eigenstate of the instantaneous Hamiltonian; they satisfy the time-independent discrete 
Schr$\ddot{\mbox{o}}$dinger equation
\begin{eqnarray}
&&\sum_l\hbar\omega_{m,l}(R) \phi_n(l,R) \nonumber\\
&&+ V_0(m,R)\phi_n(m,R) = E_n(R)\phi_n(m,R).
\label{SE2}
\end{eqnarray}
We seek the transition rates and potential that generates 
$\phi_n(m,R_f)\exp[-(i/\hbar)\int_0^{T_F}E_n(R(t'))dt']$ from $\phi_n(m,R_i)$, 
where $R(T_F)=R_f$.  
Although such dynamics is realized as a solution of Eq. (\ref{SE1}) if $dR(t)/dt$ is 
sufficiently small, corresponding to an adiabatic process, if $dR(t)/dt$ is not 
very small unwanted excitations occur.  We consider a time-dependent 
intermediate state wave function $\Psi_{\rm{FF}}$ that evolves 
from $\phi_n(m,R_i)$ to 
$\phi_n(m,R_f)\exp[-(i/\hbar)\int_0^{T_F}E_n(R(t'))dt']$ in time $T_F$.  
The Schr$\ddot{\mbox{o}}$dinger equation for $\Psi_{\rm FF}$ is
\begin{eqnarray}
i\frac{d\Psi_{\rm FF}(m,t)}{dt} &=& \sum_l\omega^{\rm FF}_{m,l}(t)\Psi_{\rm FF}(l,t)
\nonumber\\
&&+\frac{V_{\rm FF}(m,t)}{\hbar} \Psi_{\rm FF}(m,t),\nonumber\\
\label{SE3}
\end{eqnarray}
and the transition rates $\omega_{m,l}^{\rm FF}$ between $|l\rangle$ and $|m\rangle$
are time-dependent and/or tunable.  The wave function $\Psi_{\rm FF}(m,t)$ 
is assumed to be represented, with the additional phase $f(m,t)$, in the form
\begin{eqnarray}
\Psi_{\rm FF}(m,t) &=& \phi_n(m,R(t))
\exp[if(m,t)]\nonumber\\
&&\times\exp\Big{[} -\frac{i}{\hbar} \int_0^t E_n(R(t'))dt' \Big{]}.\nonumber\\
\label{psiff}
\end{eqnarray}
We require that $f(m,0)=f(m,T_F)=0$.  Assuming $\Psi_{\rm FF}(m,t)\ne 0$ 
($\phi_n(m,R(t))\ne 0$) we divide Eq. (\ref{SE3}) by $\Psi_{\rm FF}(m,t)$, 
substitute into Eq. (\ref{psiff}), and then decompose the equation into real 
and imaginary parts.  The imaginary part of the equation leads to
\begin{eqnarray}
&&\frac{dR}{dt}\mbox{Re}\Big{[}\phi_n^\ast(m,R)
\frac{\pa \phi_n(m,R)}{\pa R} \Big{]}  \nonumber\\
&&=\sum_l\mbox{Im} \Big{[} \phi_n^\ast(m,R)\phi_n(l,R)\nonumber\\
&&\times\Big{(} \omega_{m,l}^{\rm FF}(t)\exp\big{[}
i\big{(}  f(l,t)-f(m,t) \big{)} \big{]}\nonumber\\
&&-\omega_{m,l}\big{(}R(t)\big{)} \Big{)}\Big{]}
\label{SE4}
\end{eqnarray}
and the real part leads to the driving potential
\begin{eqnarray}
&&V_{\rm FF}(m,t) = V_0(m,R(t)) \nonumber\\
&& + \sum_l \mbox{Re}\Big{[}
\hbar\frac{\phi_n(l,R(t))}{\phi_n(m,R(t))}
\Big{(} \omega_{m,l}(R(t))-\omega_{m,l}^{\rm FF}(t)\nonumber\\
&& \times\exp\big{[} 
i\big{(} f(l,t) - f(m,t) \big{)} \big{]} \Big{)} \Big{]} -\hbar\frac{df(m,t)}{dt} \nonumber\\
&&- \hbar\frac{dR}{dt}
\mbox{Im}\Big{[} \frac{1}{\phi_n(m,R(t))}\frac{\pa\phi_n(m,R(t))}{\pa R}\Big{]}.
\label{VFF1}
\end{eqnarray}
When $\phi_n(m,R)=0$ for any $R$, the Schr$\ddot{\mbox{o}}$dinger 
equation (\ref{SE3}) 
takes the form
\begin{eqnarray}
\sum_l\omega_{m,l}^{\rm FF}(t) e^{if_l}\phi_n(l,R(t)) = 0,
\label{SE5}
\end{eqnarray}
and Eq. (\ref{SE2}) becomes
\begin{eqnarray}
\sum_l\omega_{m,l}(R(t))\phi_n(l,R(t)) = 0.
\label{SE6}
\end{eqnarray}
If $\phi_n(m,R(t))=0$ for any $t$ the driving potential is arbitrary because it has no influence in the Schr$\ddot{\mbox{o}}$dinger equation.

\subsection{Application to a STIRAP Process}
In its simplest form STIRAP is used to transfer population between 
states $|1\rangle$ and $|3\rangle$ in a three state manifold 
in which transitions $|1\rangle\rightarrow|2\rangle$ and 
$|2\rangle\rightarrow|3\rangle$ are 
allowed but $|1\rangle\rightarrow|3\rangle$ is forbidden.  
The driving optical field consists of two 
suitably timed and overlapping laser pulses with the (Stokes) pulse 
driving the $|2\rangle\rightarrow|3\rangle$ transition preceding 
the (pump) pulse driving the $|1\rangle\rightarrow|2\rangle$ transition.  
The field dressed states of this system are combinations of the bare 
states $|1\rangle$ and $|3\rangle$ with coefficients 
that depend on the Rabi frequencies of the 
pump ($\Omega_p$) and Stokes ($\Omega_S$) fields.  
Consequently, as those fields vary 
in time there is an adiabatic transfer of population from 
$|1\rangle$ to $|3\rangle$.  
In the three-state system the efficiency of STIRAP is relatively 
insensitive to the details of the pulse profile and the pulse 
separation  \cite{Bergmann} when the adiabatic condition 
$\Delta T(\Omega_S^2 +\Omega_p^2)^{1/2}>10$ can be met, 
where $\Delta T$ is the pulse overlap.  Using the interaction representation 
and the rotating wave approximation (RWA), the Hamiltonian of the 
three-state system with resonant pump $|1\rangle\rightarrow|2\rangle$ and 
Stokes $|2\rangle\rightarrow|3\rangle$ fields can be 
represented in the form
\begin{eqnarray}
H_{\rm RWA}(t) = -\hbar\left( \begin{array}{ccc}
0 & \Omega_p(t) & 0 \\
\Omega_p(t) & 0 & \Omega_S(t) \\
0 & \Omega_S(t) & 0
\end{array}
 \right),
\label{Hami1}
\end{eqnarray}
with $\Omega_p$ and $\Omega_S$ the Rabi frequencies defined by
\begin{eqnarray}
\Omega_p(t) = \mu_{12}E_p^{(e)}(t)/(2\hbar),\nonumber\\
\Omega_S(t) = \mu_{23}E_S^{(e)}(t)/(2\hbar),
\end{eqnarray}
where $E_{p(S)}^{(e)}$ is the envelope of the amplitude of the 
pump (Stokes) field 
and $\mu_{ij}$ the transition dipole moment between states $|i\rangle$ and 
$|j\rangle$.  
Note that, by assumption, $\mu_{13}=0$.  The time-dependent field-dressed 
eigenstates of this system are linear combinations of the 
field-free states with coefficients that depend on the Stokes 
and pump field magnitudes and the transition dipole moments. 
The field-dressed state of interest to us is
\begin{eqnarray}
|\phi_2(t)\rangle = \cos\Theta(t)|1\rangle - \sin\Theta(t)|3\rangle,
\end{eqnarray}
where
\begin{eqnarray}
\tan\Theta(t) = \frac{\Omega_p(t)}{\Omega_S(t)}.
\end{eqnarray}
Because the Stokes pulse is applied before 
but overlaps the pump pulse, 
initially $\Omega_p \ll \Omega_S$ and all of the population is 
initially in field-free state $|1\rangle$. 
 At the final time $\Omega_p \gg \Omega_S$ so all of the population 
in $|\phi_2(t)\rangle$ projects onto the target 
state $|3\rangle$.  
Note that $|\phi_2(t)\rangle$ has no projection on the 
intermediate field-free state $|2\rangle$.  
Suppose now that either the pulsed field duration or the field strength 
must be restricted to avoid exciting unwanted processes 
that compete 
with the desired population transfer, with the consequence that the 
condition $\Delta T(\Omega_S^2+\Omega_p^2)^{1/2}>10$ cannot be met.  
Then the STIRAP process generates incomplete 
population transfer and we propose to assist the population transfer 
with a fast-forward driving field.

The analysis of the preceding subsection can be applied to a 
three-state STIRAP process with $V_0=0$ and the identifications 
$\omega_{1,3}=0$, $\omega_{1,2}(R(t))=\Omega_p(R(t))$, 
$\omega_{2,3}(R(t)) = \Omega_S(R(t))$.
We choose $R(t) = t$, in which case $\omega_{1,2}$ and $\omega_{2,3}$ 
correspond to the Rabi frequencies of the pump and Stokes pulses.  
The Hamiltonian corresponding to the time-independent 
Schr$\ddot{\mbox{o}}$dinger 
equation (\ref{SE2}) is represented as Eq. (\ref{Hami1}).  We now consider a field-dressed state
\begin{eqnarray}
|\phi_2(R)\rangle = \sum_m \phi_2(m,R) |m\rangle
\end{eqnarray}
with $\phi_2(1,R)=\cos\Theta(R)$, $\phi_2(2,R)=0$, 
$\phi_2(3,R)=-\sin\Theta(R)$, and 
$\phi_2(3,R)/\phi_2(1,R)=-\Omega_p(R)/\Omega_S(R)$.
As mentioned earlier, $m=2$ is treated separately because 
$\phi_2(2,R)=0$.  For $m=2$ Eqs. (\ref{SE5}) and (\ref{SE6}) take the form
\begin{eqnarray}
&&\omega_{2,1}^{\rm FF}(t) e^{if_1}\phi_2(1,R) \nonumber\\
&&+ \omega_{2,3}^{\rm FF}(t)e^{if_3}\phi_2(3,R) = 0,
\label{SE7}
\end{eqnarray}
and 
\begin{eqnarray}
\omega_{2,1}(R) \phi_2(1,R) 
+ \omega_{2,3}(R) \phi_2(3,R) = 0,\nonumber\\
\label{SE8}
\end{eqnarray}
respectively.  Combining Eqs. (\ref{SE7}) and (\ref{SE8}) we obtain
\begin{eqnarray}
\frac{\omega_{2,1}^{\rm FF}(t)}{\omega_{2,3}^{\rm FF}(t)}
=e^{i\Delta f}\frac{\omega_{2,1}(R(t))}{\omega_{2,3}(R(t))},
\label{omegaff}
\end{eqnarray}
with
\begin{eqnarray}
\Delta f(t) \equiv f_3(t) - f_1(t).
\end{eqnarray}
Noting that $\phi_2(2,R)=0$ and $\phi_2(1,R), \phi_2(3,R)\in R$,
Eq. (\ref{SE4}) can be rewritten as 
\begin{eqnarray}
\frac{dR(t)}{dt}\frac{\pa \phi_2(1,R)}{\pa R}
=\phi_2(3,R) \mbox{Im} \Big{[} \omega_{1,3}^{\rm FF}(t) e^{i\Delta f}\Big{]},
\nonumber\\
\frac{dR(t)}{dt}\frac{\pa \phi_2(3,R)}{\pa R}
=\phi_2(1,R) \mbox{Im} \Big{[} \omega_{3,1}^{\rm FF}(t) e^{-i\Delta f}\Big{]}.
\nonumber\\
\label{SE9}
\end{eqnarray}
It can be shown that the two equations (\ref{SE9}) are 
identical by using the relations $(\omega_{3,1}^{\rm FF})^\ast = \omega_{1,3}^{\rm FF}$
and $\pa_R\phi_2(1,R) / \phi_2(3,R) = -\pa_R\phi_2(3,R) / \phi_2(1,R)$, 
which are directly derived from 
$\pa_R\big{(} |\phi_2(1,R)|^2 + |\phi_2(3,R)| \big{)}^2=0$.  
Equations (\ref{omegaff}) and (\ref{SE9}) determine the Rabi frequencies.

\begin{figure}[h!]
\begin{center}
\includegraphics[width=8cm]{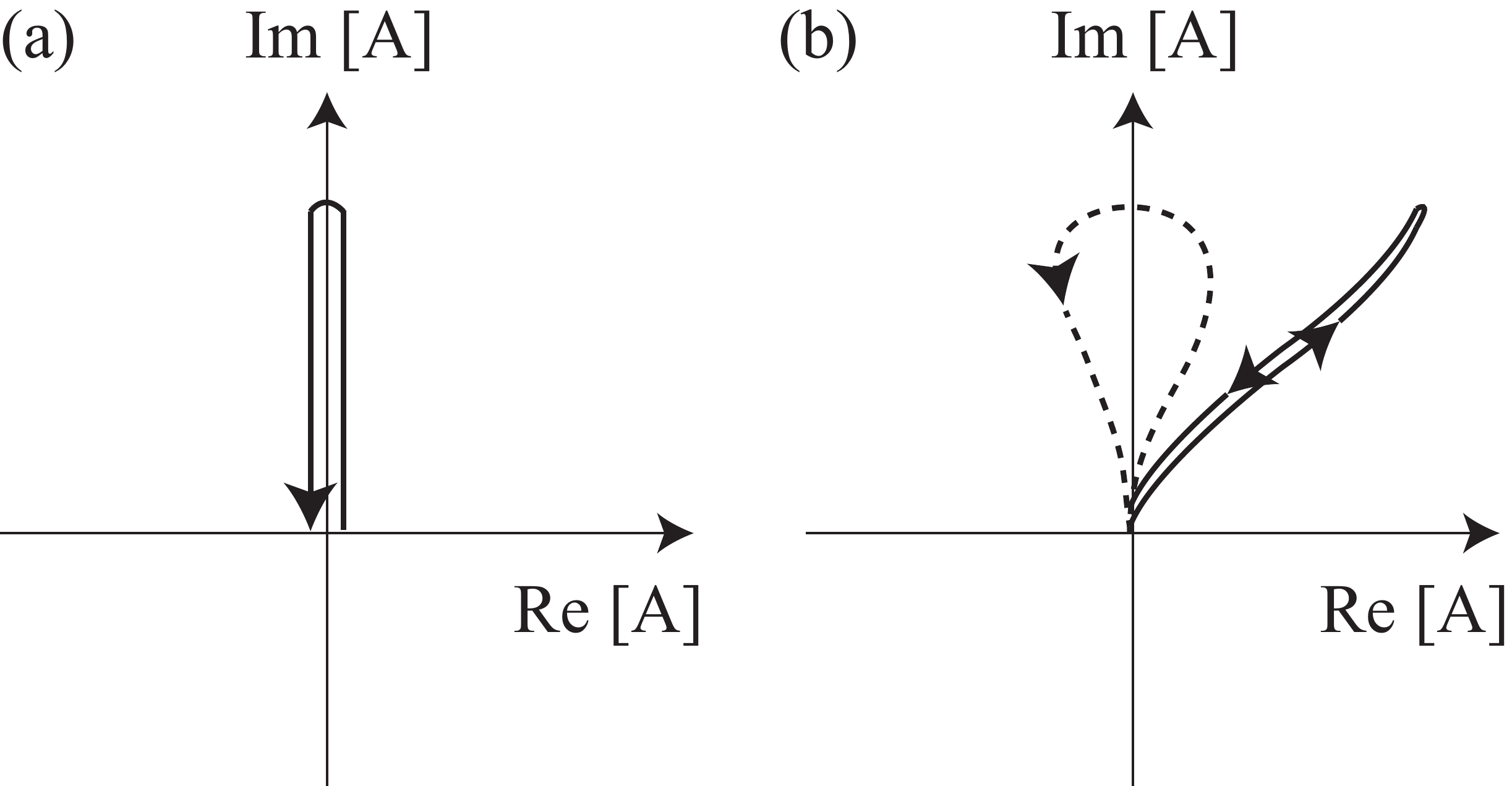}
\end{center}
\caption{\label{fig:epsart} 
Schematic diagrams of three different trajectories of $A$: 
(a) $\mbox{Re}[A(t)]=0$; (b) $\mbox{Re}[A(t)]=A_1\exp [-t^2 / \tau^2]$ 
(solid curve) and $\mbox{Re}[A(t)]=A_2 (t/\tau) \exp [-t^2 / \tau^2]$ 
(dotted curve) with $A_{1(2)}$ and $\tau$ constants.}
\label{vecA}
\end{figure}

We consider a fast-forwarded STIRAP process with finite $f_m$ and 
vanishing diagonal elements of the driving Hamiltonian, 
$V_{\rm FF}=0$.  Equation (\ref{VFF1}) and $V_{\rm FF}=0$ lead to
\begin{eqnarray}
\frac{\omega_{1,2}}{\omega_{2,3}}\mbox{Re}\big{[} A(t) \big{]}
-\frac{df_1}{dt} = 0,\nonumber\\
\frac{\omega_{2,3}}{\omega_{1,2}}\mbox{Re}\big{[} A(t) \big{]}
-\frac{df_3}{dt} = 0,
\end{eqnarray}
with
\begin{eqnarray}
A(t) = \omega_{1,3}^{\rm FF}(t) e^{i\Delta f(t).}
\end{eqnarray}
$d(\Delta f)/dt$ is determined when we choose $\mbox{Re}[A(t)]$ to be
\begin{eqnarray}
\frac{d\Delta f(t)}{dt} = \Big{(} \frac{1}{\tan\Theta(t)} 
- \tan\Theta(t) \Big{)} \mbox{Re}\big{[} A(t) \big{]}.\nonumber\\
\end{eqnarray}
Equation (\ref{SE9}) determines the imaginary part of $A$ to be
\begin{eqnarray}
\mbox{Im} \big{[} A(t) \big{]} = \frac{d\Theta}{dt}.
\end{eqnarray}
Then $\omega_{1,3}^{\rm FF}$ is represented as
\begin{eqnarray}
\omega_{1,3}^{\rm FF} &=& e^{-i\Delta f} \Big{(} \mbox{Re} [A] 
+ i\frac{d\Theta}{dt} \Big{)}\nonumber\\
&& = e^{-i\Delta f} \Big{[} \frac{\sin 2\Theta}{2\cos2\Theta}
\frac{d\Delta f}{dt} + i\frac{d\Theta}{dt} \Big{]}.\nonumber\\
\label{omegaff2}
\end{eqnarray}
The Rabi frequency $\omega_{1,3}^{\rm FF}(t)$ is complex, and must be 
realized by controlling the time dependences of the phases of the 
laser fields as well as the relative phase between and $\omega_{1,2}^{\rm FF}$
and $\omega_{2,3}^{\rm FF}$.  
There is an arbitrariness in the choice of $\mbox{Re}[A(t)]$ or 
$\Delta f(t)$. Three different 
trajectories of $A$ are depicted schematically in Fig. \ref{vecA}, for 
$\mbox{Re}[A(t)]=0$, $\mbox{Re}[A(t)]=A_1\exp [-t^2 / \tau^2]$
and $\mbox{Re}[A(t)]=A_2 (t/\tau) \exp [-t^2 / \tau^2]$, with $A_{1(2)}$
and $\tau$ constant.

It can be shown that the fast-forward assisted STIRAP protocol gives 
the same Rabi frequencies as does the counter-diabatic field assisted 
STIRAP protocol with 
\begin{eqnarray}
f_m=0.
\end{eqnarray}
Equations (\ref{omegaff}) and (\ref{omegaff2}) lead to 
\begin{eqnarray}
\frac{\omega_{2,1}^{\rm FF}(t)}{\omega_{2,3}^{\rm FF}(t)} = 
\frac{\omega_{2,1}(R(t))}{\omega_{2,3}(R(t))},
\label{o_ratio1}
\end{eqnarray}
and 
\begin{eqnarray}
\omega_{1,3}^{\rm FF} = i\frac{d\Theta}{dt}.
\label{omegaff3}
\end{eqnarray}
$\omega_{1,3}^{\rm FF}$ in Eq. (\ref{omegaff3}) is the same as the Rabi frequency of the CDF.  
The trajectory of $A$ with $\mbox{Re}[A(t)]=0$ depicted in Fig. \ref{vecA}(a) corresponds to the CDF.  
Equation (\ref{o_ratio1}) determines the ratio of $\omega_{2,1}^{\rm FF}(t)$ and 
$\omega_{2,3}^{\rm FF}(t)$ but their intensities 
are arbitrary and can even be zero, consistent with the observation that 
the CDF alone can generate complete population transfer in a two-level system.
When $\mbox{Re}[A(t)]\ne 0$ 
the pulse area of the FFF pulse is larger than $\pi$, 
in contrast to the pulse area of the CDF, which is 
$\pi$ \cite{Demirplak1, Chen}. The restriction of the pulse area that is characteristic 
of the CDF protocol is eased in the fast-forward protocol.  

\section{HCN$\rightarrow$HNC Isomerization Reaction}
\label{HCN}
Previous studies of STIRAP generated population transfer in 
laser-assisted HCN $\rightarrow$ HNC isomerization 
\cite{Cheng, Jakubetz} have revealed 
that background states coupled to the 
subset of states used by the driving 
STIRAP process degrade the population transfer efficiency.  
Mitigation of this inefficiency is sought in an assisted STIRAP process.

\begin{figure}[h!]
\begin{center}
\includegraphics[width=8cm]{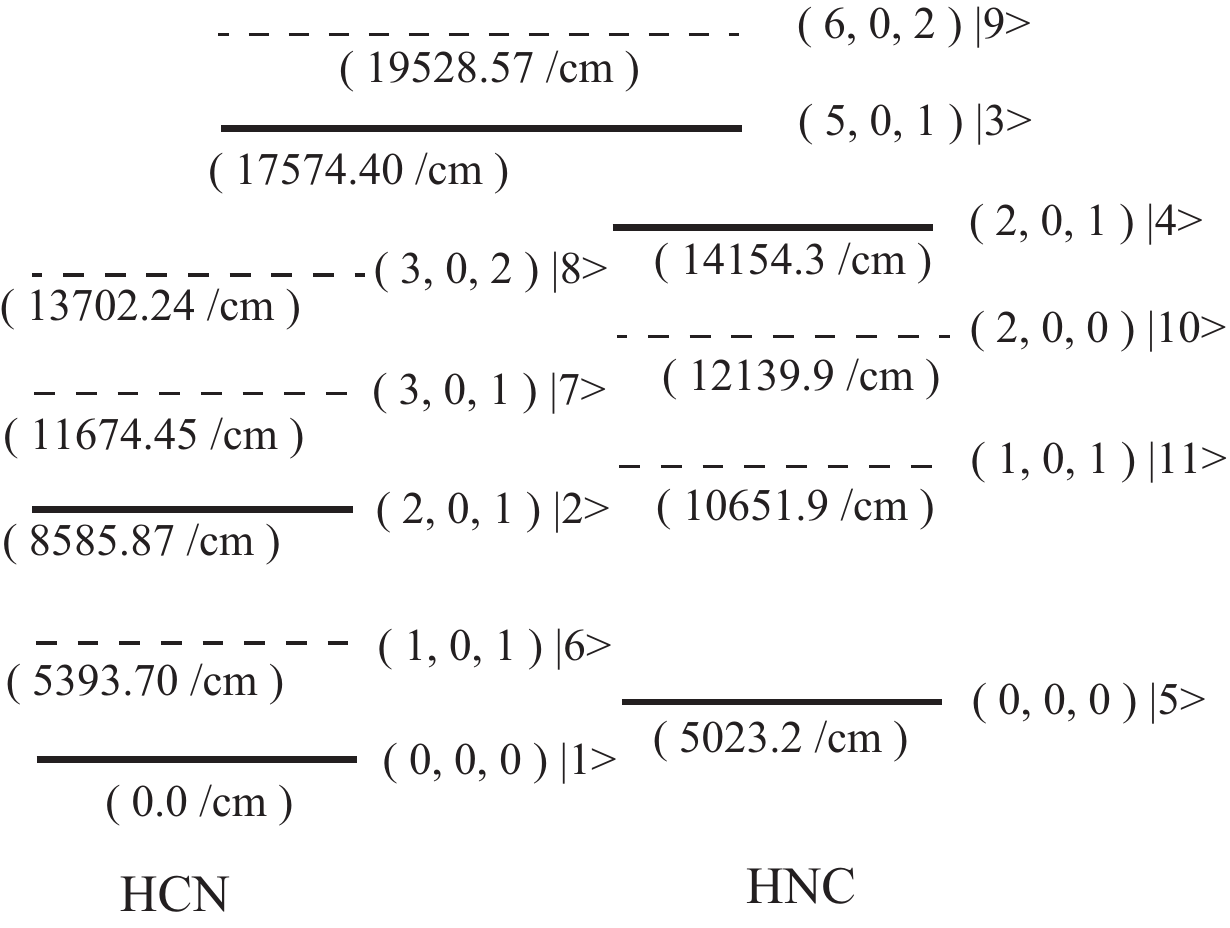}
\end{center}
\caption{\label{fig:epsart} 
Schematic diagram of the vibrational spectrum of states used 
for the numerical simulations.  The states selected for use in the successive 
STIRAP processes are represented with thick lines, and the background 
states are represented with thin dashed lines.}
\label{lavels_HCN}
\end{figure}

The three-dimensional potential energy surface for non-rotating HCN/HNC 
has been well studied \cite{Smith, Yang, Jonas, Bowman, Jakubetz2}.  
The key degrees of freedom that characterize this surface are the CH, NH 
and CN stretching motions and the CNH bending motion. These are combined 
in the symmetric stretching, bending and asymmetric stretching normal modes, 
with quantum numbers $(\nu_1, \nu_2, \nu_3)$, respectively.  
The vibrational energy 
levels of HCN and HNC have been calculated by Bowman et al \cite{Bowman}.  
Driving the ground state-to-ground state
HCN $\rightarrow$ CNH isomerization with a conventional 
STIRAP process that uses two monochromatic laser fields is difficult 
because the Franck-Condon factors between the ground vibrational states 
$(0, 0, 0)$ of HCN and CNH and the vibrational levels close to the top of 
the isomerization barrier (e.g. $(5, 0, 1)$) are extremely small.  
In the model system considered by Kurkal and Rice \cite{Kurkal2} eleven 
vibrational states, shown schematically in Fig. \ref{lavels_HCN}, 
are considered; 
rotation of the molecule is neglected.  Kurkal and Rice proposed 
overcoming the Franck-Condon barrier with sequential STIRAP, consisting 
of two successive STIRAP processes.  The use of this sequence is 
intended to avert the unwanted competition with other processes that can 
be generated by the very strong fields that would be needed to overcome the 
Franck-Condon barriers encountered in a single STIRAP process.  
In the first step of the sequential STIRAP, Kurkal 
and Rice chose the $(0, 0, 0)$, $(2, 0, 1)$ and $(5, 0, 1)$ 
states of HCN as the 
initial, intermediate and final states, respectively; in the second STIRAP 
process, the $(5,0,1)$, $(2,0,1)$ and $(0,0,0)$ states of HNC are taken as the 
initial, intermediate and final states, respectively.  Other states, shown 
with dashed lines in Fig. \ref{lavels_HCN}, 
are regarded as background states. 
The pump 1 field is resonant with the transition from the (0,0,0) 
state of HCN to the $(2,0,1)$ state of HCN; the Stokes 1 field is 
resonant with the transition from the $(2,0,1)$ state of HCN to the $(5,0,1)$ 
state; the pump 2 field is resonant with the transition from the $(5,0,1)$ 
state to the $(2,0,1)$ state of HNC; and the Stokes 2 field is resonant with 
the transition from the $(2,0,1)$ state to the $(0,0,0)$ state of HNC.  
The transition dipole moments and the energies of the vibrational 
states denoted by $|i\rangle$ are listed in Ref. 12.

\begin{figure}[h!]
\begin{center}
\includegraphics[width=8cm]{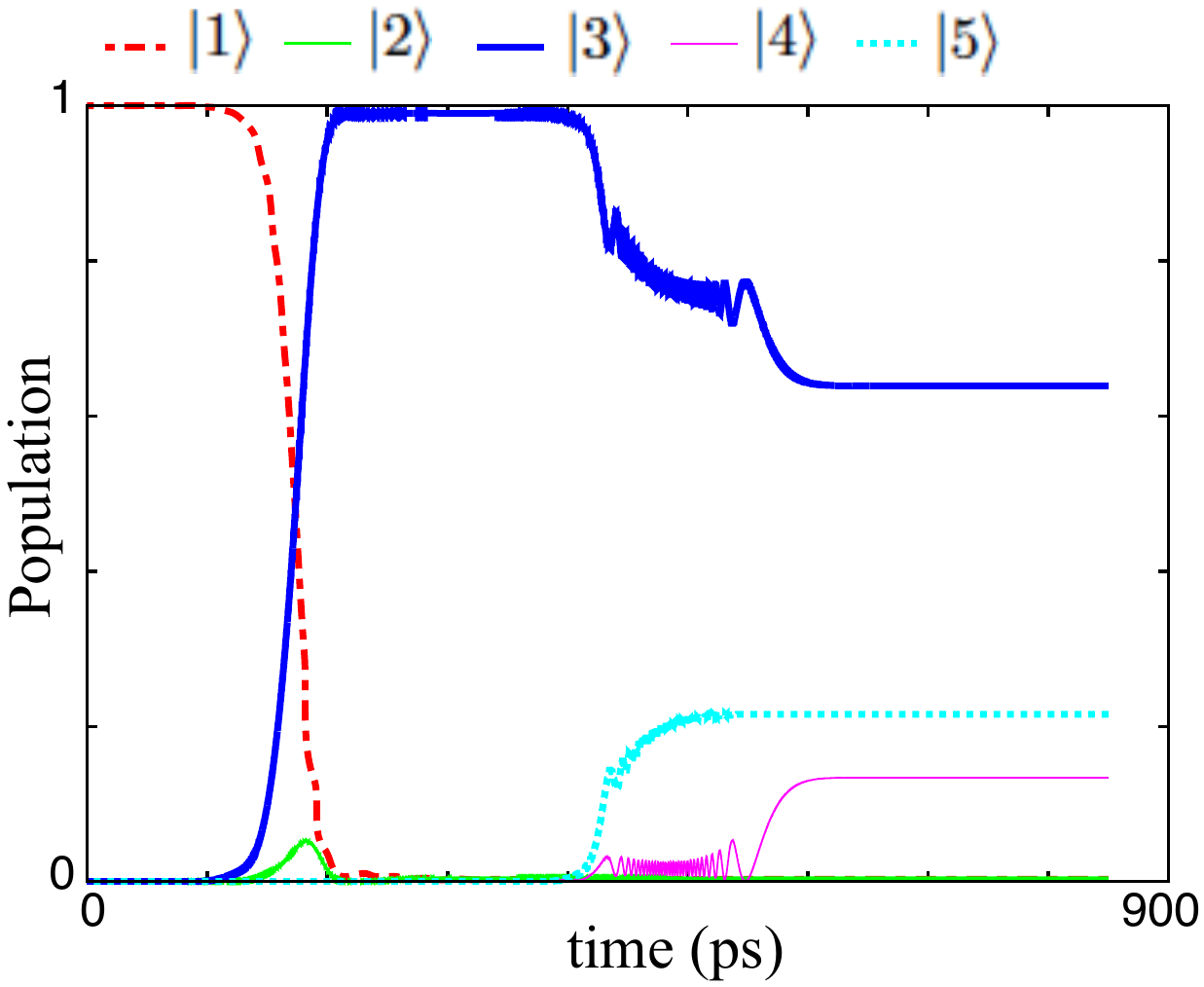}
\end{center}
\caption{Time-dependence of the several state populations for $|1\rangle 
\rightarrow|5\rangle$ in the sequential STIRAP driven HCN $\rightarrow$ 
HNC isomerization \cite{Masuda-Rice2}.}
\label{sequential_STIRAP}
\end{figure}

Kurkal and Rice showed that the first STIRAP process is not sensitive 
to coupling with the background states caused by the Stokes 1 pulse 
\cite{Kurkal2}.  However the second STIRAP process is influenced by 
interference with the background states because the intermediate state 
of the second STIRAP process has large transition dipole moments 
with the background states \cite{Masuda-Rice2}.  The time-dependences of 
the populations of states $|1\rangle-|5\rangle$ in the sequential 
STIRAP process are displayed in Fig. \ref{sequential_STIRAP}.  
We take the strengths of the pump and the Stokes fields to be
\begin{eqnarray}
E_{j,p(S)}^{(e)}(t) = \tilde{E}_{j,p(S)} 
\exp\Big{[} -\frac{(t-T_{j,p(S)})^2}{(\Delta \tau)^2} \Big{]}\nonumber\\
\end{eqnarray}
where $\Delta \tau = \mbox{FWHM}/(2\sqrt{\ln 2})$, and FWHM is the full width at half maximum of the Gaussian 
pulse with maximum intensity $\tilde{E}_{j,p(S)}$ 
that is centered at $T_{j,p(S)}$, 
and $j=(1,2)$ denotes the 
first ($j = 1$) and the second STIRAP ($j = 2$) process.  We solved the 
time-dependent Schr$\ddot{\mbox{o}}$dinger equation numerically with a fourth order 
Runge-Kutta integrator in a basis of bare matter eigenstates with 
$T_{j,p}-T_{j,S} = \mbox{FWHM}/(2\sqrt{\ln 2})$. 
The parameters of the laser fields used in our calculations are shown 
in Table \ref{table_laser_seq}.  It is seen clearly in Fig. \ref{sequential_STIRAP} that the fidelity of the first STIRAP in the 
sequential STIRAP process is robust with respect to interference from 
the background states \cite{Masuda-Rice2}.  For that reason we assume that 
the population is transferred from $|1\rangle$ to $|3\rangle$ 
completely, and we focus attention 
on the second STIRAP process, choosing $|3\rangle$ as the initial state of the assisted 
STIRAP control process.  States $|1\rangle$ and $|2\rangle$ 
are now and hereafter regarded 
as background states.
\begin{table}[htb]
  \caption{Strengths and widths of the pump 1, 2 and Stokes 1, 2 laser pulses.
}
  \begin{tabular}{|l|c|c|c|}
    \hline
         & $\tilde{E}_{j,p(S)}$ (a.u.)  & $T_{j,p(S)}$ & FWHM (ps) \\
    \hline
     Stokes 1   & 0.00692 & 133 & 85 \\
     pump 1 & 0.00728 & 194 & 85 \\
     Stokes 2   & 0.00575 & 423 & 85 \\
     pump 2 & 0.00220 & 484 & 85 \\
\hline
  \end{tabular}
\label{table_laser_seq}
\end{table}

\begin{figure}[h!]
\begin{center}
\includegraphics[width=8cm]{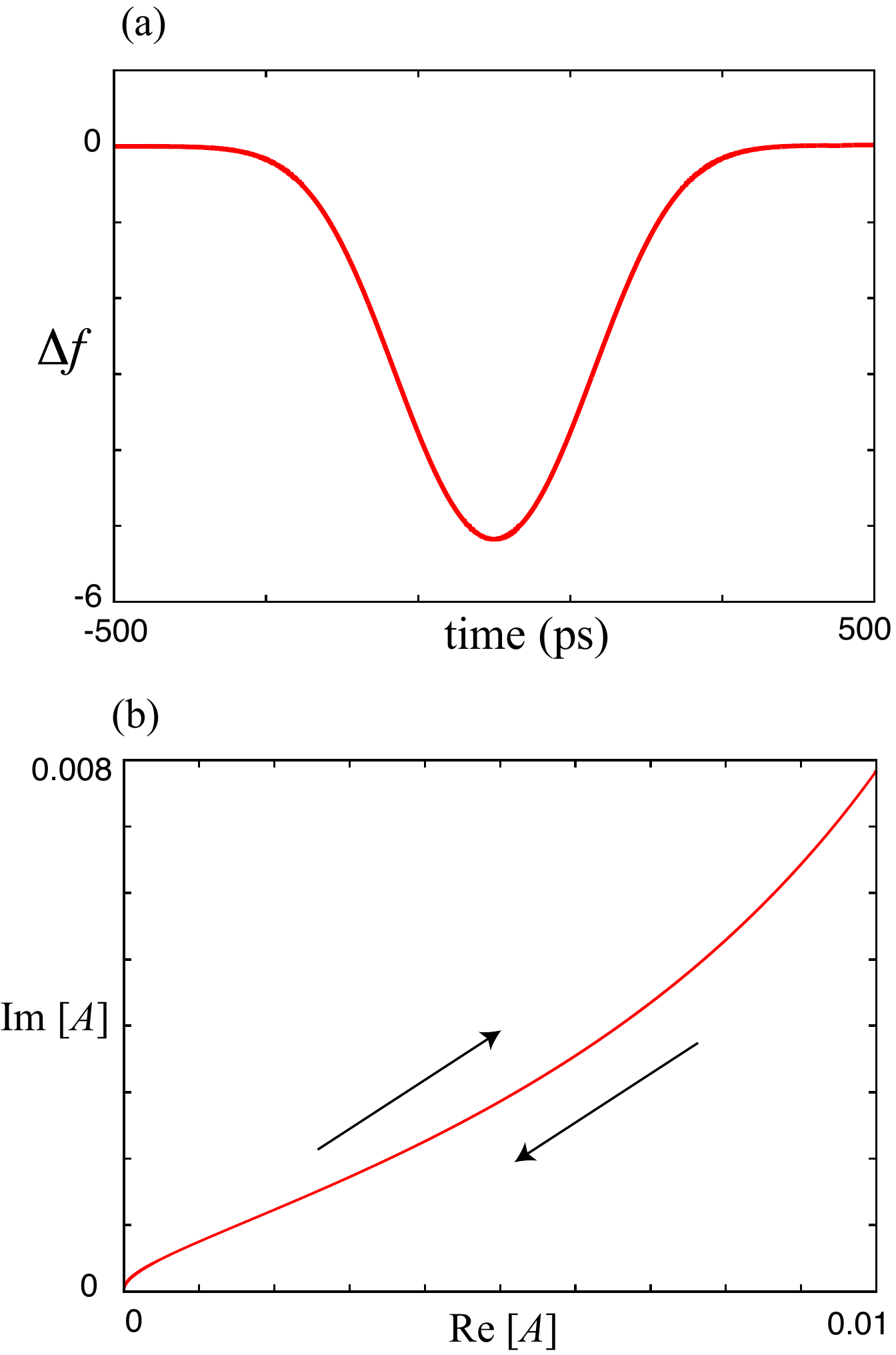}
\end{center}
\caption{The time-dependence of (a) $\Delta f$ and (b) $A$ 
for FWHM = 212.5 ps and $T_{2,p}-T_{2,S}=\mbox{FWHM}/(2\sqrt{\ln 2})$.}
\label{fa_dtheta_deltaf_v}
\end{figure}

\begin{figure}[h!]
\begin{center}
\includegraphics[width=8cm]{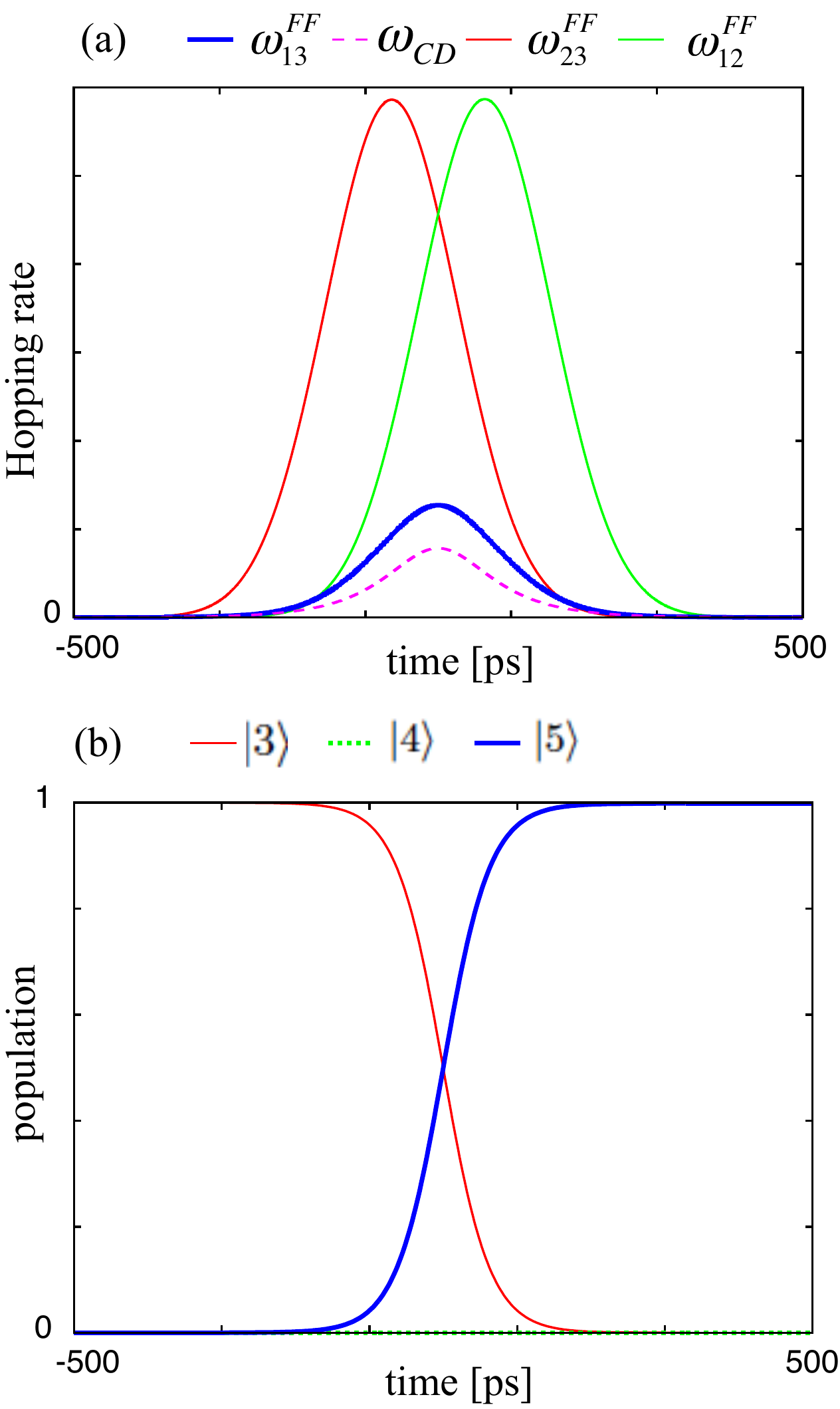}
\end{center}
\caption{ (a) The time-dependences of the amplitudes of the Rabi frequencies for FWHM = 212.5 ps and $T_{2,p}-T_{2,S}=\mbox{FWHM}/(2\sqrt{\ln 2})$. 
(b) The time-dependences of the populations.
}
\label{p_3L}
\end{figure}

We now take the three vibrational states $|3\rangle$, 
$|4\rangle$, $|5\rangle$ in Fig. \ref{lavels_HCN} as the initial, 
intermediate and target states of  
both a STIRAP + FFF and a STIRAP + CDF process.  
We use the amplitudes and FWHMs for the pump and Stokes laser 
pulses listed in Table \ref{table_laser}, and choose the time-dependence of 
$\mbox{Re}[A(t)]$ to be
\begin{eqnarray}
\mbox{Re}[A(t)] = A_0\exp\Big{[}-\frac{t^2}{\Delta \tau^2} \Big{]}
\end{eqnarray}
with $A_0=0.01$ /ps.
\begin{table}[htb]
  \caption{Strengths and widths of the pump 2 and Stokes 2 laser pulses.}
  \begin{tabular}{|l|c|c|}
    \hline
         & $\tilde{E}_{2,p(S)}$ (a.u.)  & FWHM (ps) \\
    \hline
     pump 2   & 0.0009295 & 212.5 \\
     Stokes 2 & 0.002875 & 212.5 \\
\hline
  \end{tabular}
\label{table_laser}
\end{table}
The time-dependence of $\Delta f$ is shown in Fig. 
\ref{fa_dtheta_deltaf_v}(a).  
The phase of the Stokes field is changed by $\Delta f$ 
(see Eq. (\ref{omegaff})) and the 
trajectory of $A(t)=\omega_{1,3}^{\rm FF}(t)e^{i\Delta f(t)}$ 
is shown in Fig. \ref{fa_dtheta_deltaf_v}(b).  The amplitudes of the Rabi 
frequencies coupling the three states are shown in Fig. \ref{p_3L}(a) and the 
time-dependence of the population of each state 
in the three-state 
system decoupled from the background states is shown in Fig. \ref{p_3L}(b).  
The data displayed clearly show that 100$\%$ population transfer is 
generated.  As seen from Eq. (\ref{omegaff2}) and Fig. \ref{p_3L}(a) the amplitude of 
$\omega_{1,3}^{\rm FF}$ is 
larger than that of the CDF. The restriction of the pulse area that 
is characteristic of the CDF protocol is eased in the fast-forward 
protocol. 

\begin{figure}[h!]
\begin{center}
\includegraphics[width=8cm]{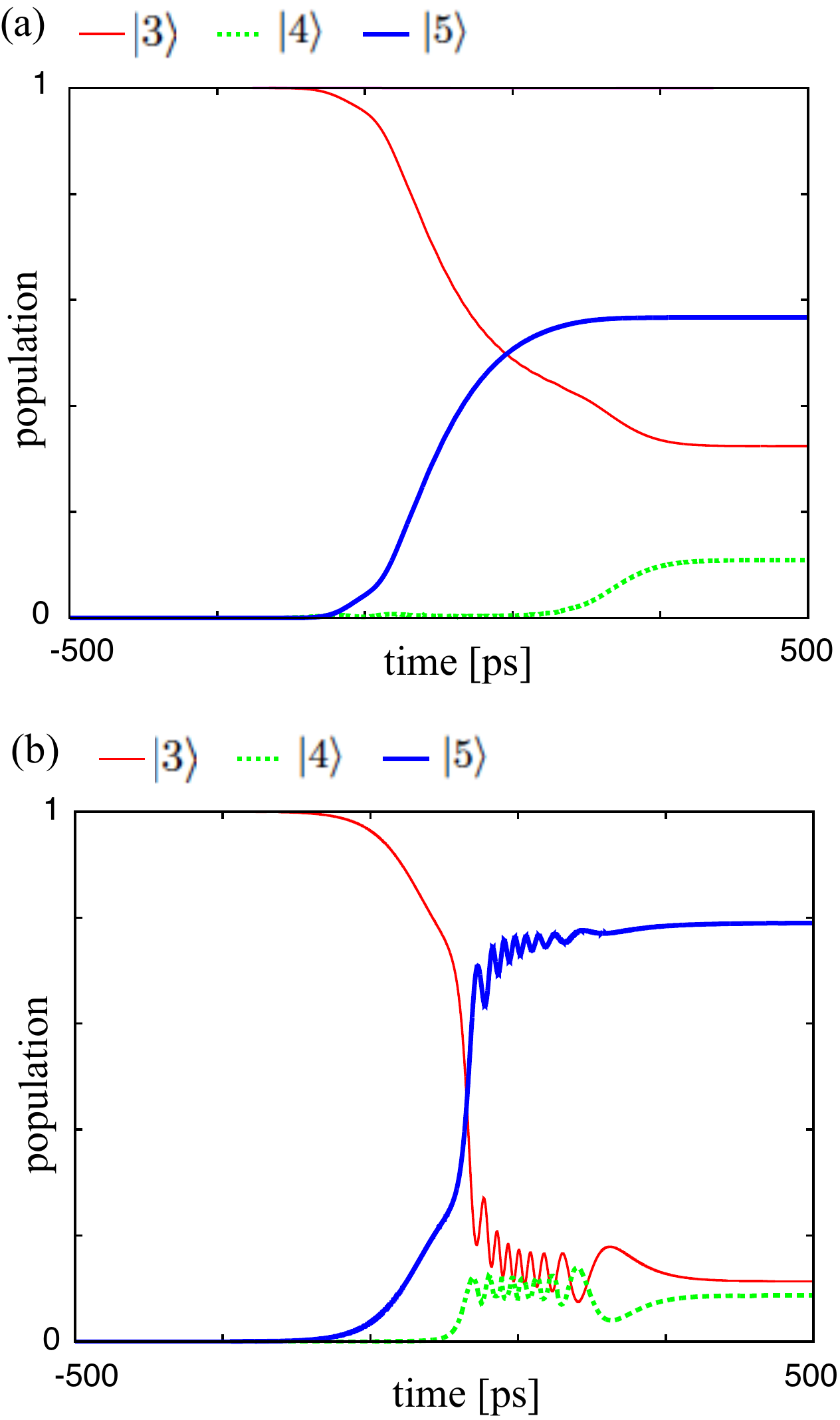}
\end{center}
\caption{ Time-dependences of the populations of the initial, 
intermediate and the target states for (a) STIRAP and (b) STIRAP + FFF 
control for peak field ratio = 1.2, FWHM = 212.5 ps and 
$T_{2,p}-T_{2,S}=\mbox{FWHM}/(2\sqrt{\ln 2})$.
}
\label{p_com_ff_1.2_v}
\end{figure}

\begin{figure}[h!]
\begin{center}
\includegraphics[width=8cm]{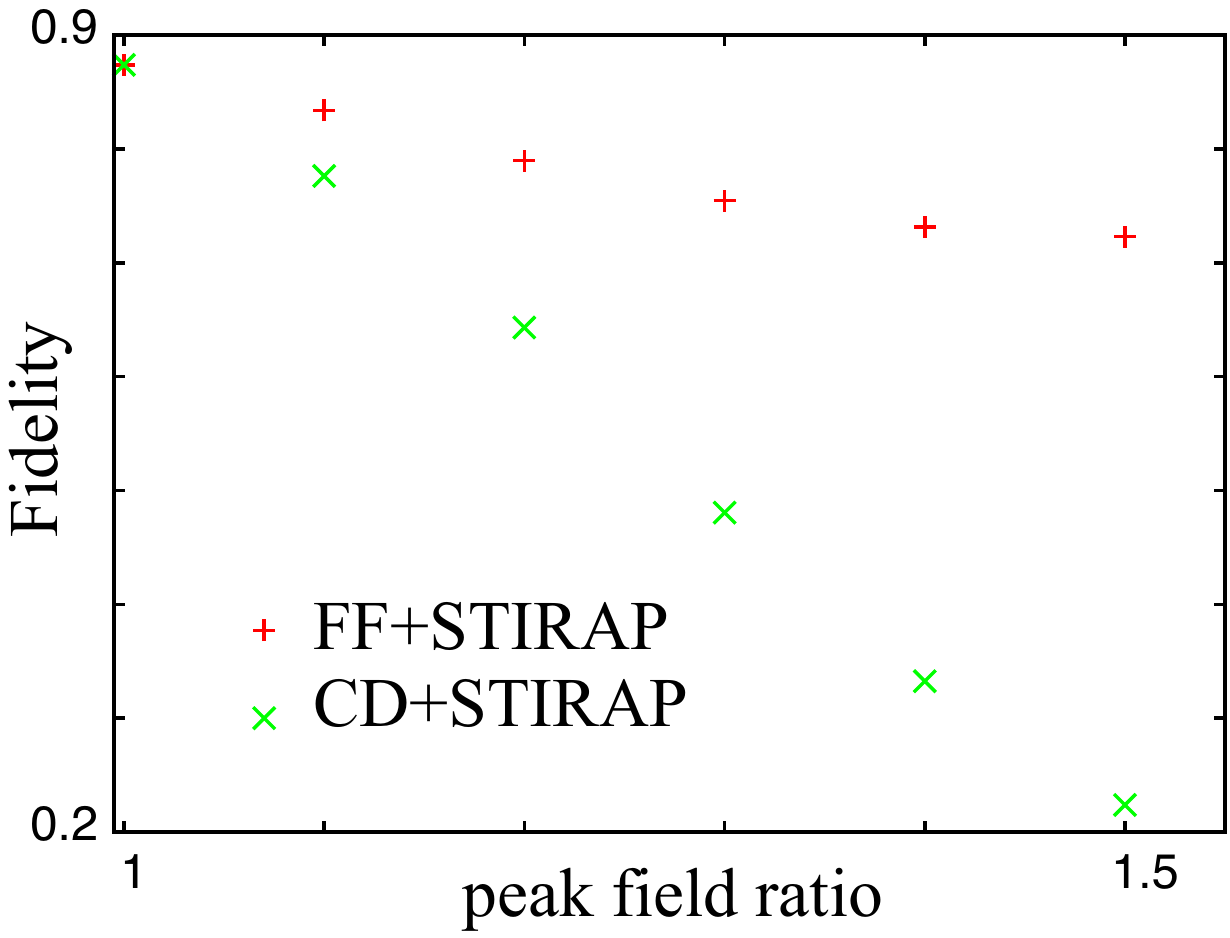}
\end{center}
\caption{
The dependence of the fidelity on the peak amplitude of the laser 
field coupling the initial and target states. The horizontal axis is 
the ratio of the peak laser field amplitude to the amplitude of the CDF 
in Eq. (\ref{omegaff3}) for FWHM = 212.5 ps and 
$T_{2,p}-T_{2,S}=\mbox{FWHM}/(2\sqrt{\ln 2})$.
}
\label{p_ff_cd_HCN}
\end{figure}

We now examine
the efficiency of the STIRAP + FFF control when the subset 
of states $|3\rangle$,$|4\rangle$,$|5\rangle$ is 
embedded in the manifold of 
states depicted in Fig. \ref{lavels_HCN}. 
We consider a FFF corresponding to Rabi frequency $\omega_{1,3}^{\rm FF}$ 
accompanied with 
a pump pulse and a phase-controlled Stokes field with all the background 
states in Fig. \ref{lavels_HCN}; 
the time-evolution of the system is calculated exactly, 
without use of the rotating wave approximation. In Fig. \ref{p_com_ff_1.2_v} 
the time-dependences 
of the populations of the initial, intermediate and the target states for 
(a) STIRAP and (b) STIRAP + FFF control for $A_0=0.0053$ 
(the peak amplitude of the 
FFF to the CDF is about 1.2) are shown for laser pulses with 
FWHM = 212.5 ps and $T_{2,p}-T_{2,S} = \mbox{FWHM}/(2\sqrt{\ln 2})$. 
The efficiencies of both STIRAP and STIRAP + FFF controls 
are degraded due to interference with background states strongly coupled to 
the intermediate state. However the influence of the background states is 
suppressed in the STIRAP + FFF control compared to that in
the STIRAP control 
because of the direct coupling of the initial and target 
states by the FFF. 
Suppose now
that the peak amplitude of the laser field coupling the 
initial and final states is larger than that of the CDF in Eq. (\ref{omegaff3}) 
and is fixed, whilst its phase and that of the Stokes field are 
controllable with respect to time.  As shown in Fig. \ref{p_ff_cd_HCN} 
the fidelity of the 
STIRAP + CDF control decreases when the amplitude of the laser field 
coupling the initial and final states is larger than that of the CDF 
in Eq. (\ref{omegaff3}).  
In such cases the decrease of the fidelity is partially 
avoidable via phase control of the laser fields. Five values of $A_0$ are used; 
the ratio of the peak amplitude of the FFF to the CDF amplitude ranges 
from 1 to 1.5.  In Fig. \ref{p_ff_cd_HCN} 
we compare the calculated fidelities to that 
of the STIRAP + CDF control for the case that the CDF strength is greater 
than that of the CDF in Eq. (\ref{omegaff3}).  
The decrease of the fidelity due to 
variance of the amplitude of the CDF is reduced by phase control of the 
laser pulses.  

\begin{figure}[h!]
\begin{center}
\includegraphics[width=8cm]{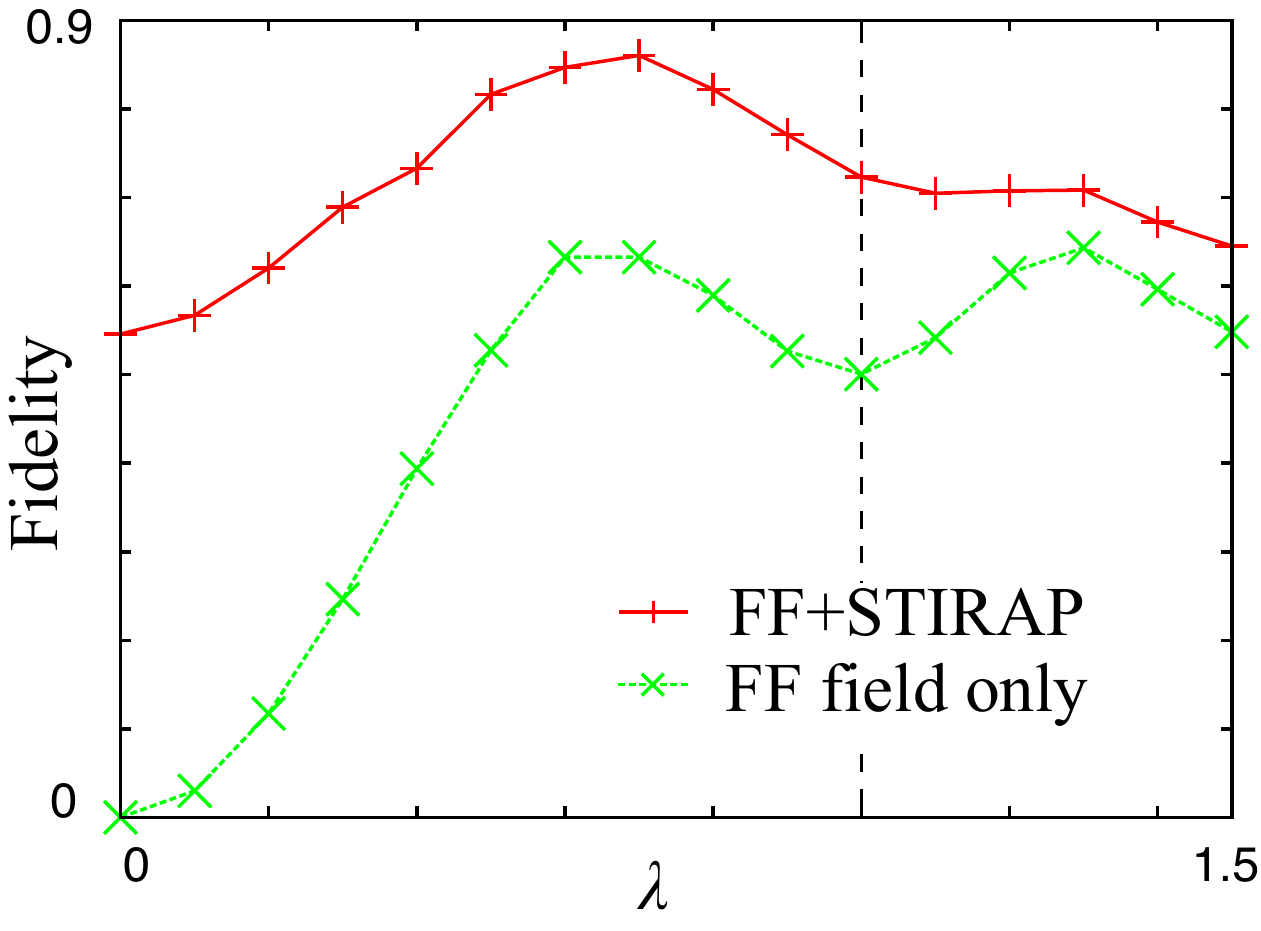}
\end{center}
\caption{
Comparison of the fidelity of STIRAP+FFF control and FFF control with FWHM = 212.5 ps and $T_{2,p}-T_{2,S}=\mbox{FWHM}/(2\sqrt{\ln 2})$ for various $\lambda$.
}
\label{fid_lam_com}
\end{figure}

As seen from Eq. (\ref{omegaff}), the FFF alone can generate complete population 
transfer to the target state in a two-level system.  However the 
efficiency of the single pulse control is degraded when there is 
interaction with the background states, and is not stable to variation 
of the area of the pulse.  
So as to study the stability of the efficiency of the population transfer driven by a variable FFF we represent the total driving field in the form
\begin{eqnarray}
E(t) = E_p(t) + E_S(t) + \lambda E_{\rm FF}(t),
\end{eqnarray}
where $\lambda = 0$ corresponds to driving the system with only the STIRAP fields and $\lambda = 1$ to driving the system with the STIRAP and the FFF;
$E_{p(S)}$ is the pump (Stokes) field; $E_{\rm FF}$ is the FFF corresponding to the Rabi frequency in Eq. (\ref{omegaff2}).
 In Fig. \ref{fid_lam_com} 
the stability of the STIRAP + FFF 
control and the FFF alone control to the variation is monitored by the 
fidelity as a function of $\lambda$.  
Clearly, the sensitivity to the variation of 
amplitude of STIRAP + FFF control is decreased compared to that of FFF 
control.  The STIRAP + FFF control generates higher fidelity than do 
STIRAP or FFF individually for a wide range of the field ratio $\lambda$.  
The value of $\lambda$ corresponding to the peak of the fidelity 
of STIRAP + FFF control is smaller than one in Fig. \ref{fid_lam_com}, because of the 
interference with the background states generated by the strong fields.

\begin{figure}[h!]
\begin{center}
\includegraphics[width=8cm]{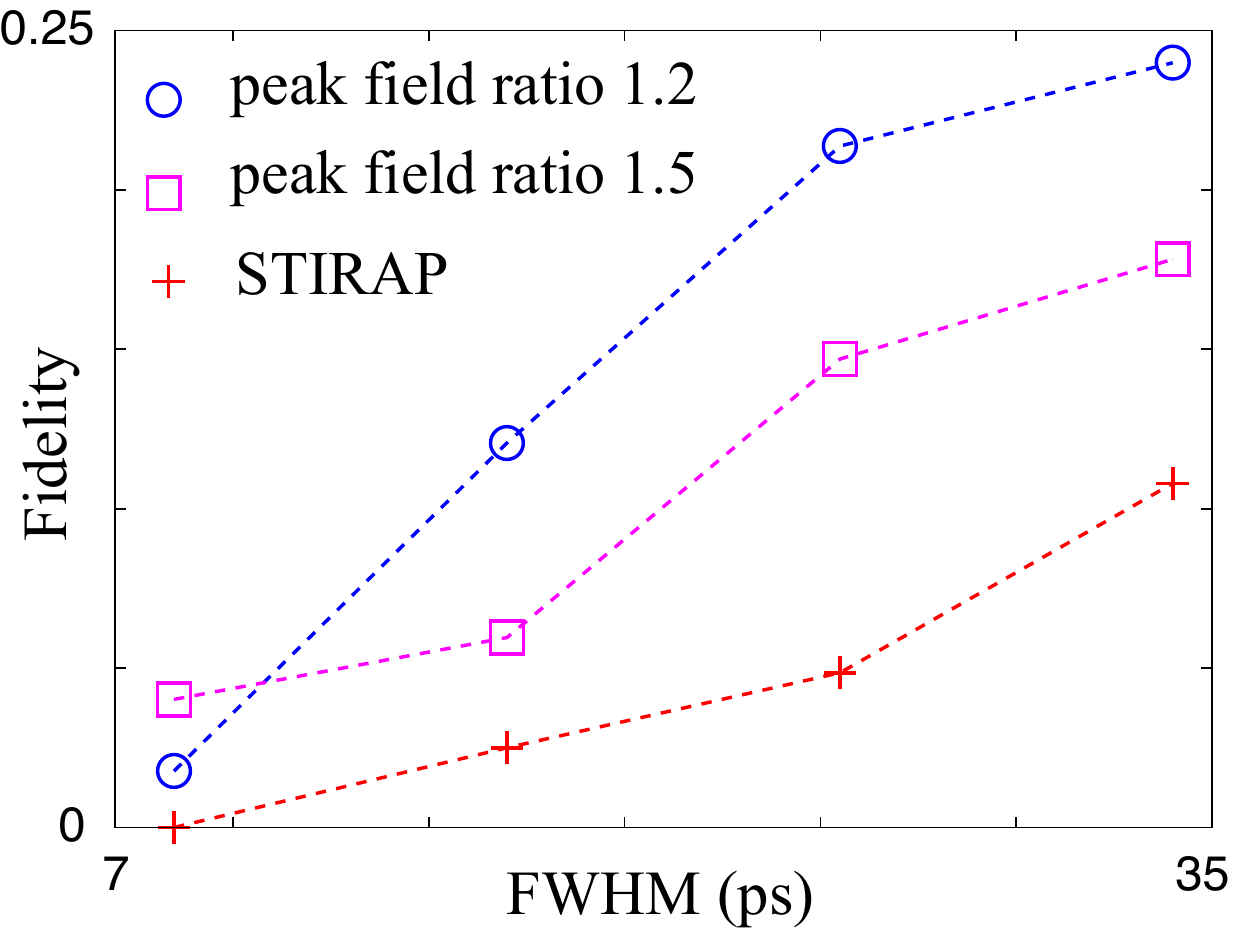}
\end{center}
\caption{
FWHM-dependence of the fidelity for the STIRAP+FFF control (FF field ratio=1.2 and 1.5), $\lambda=0.3$ and STIRAP with 
$\tilde{E}_{2,p}=0.003718$ a.u., $\tilde{E}_{2,S}=0.0115$ a.u.
and $T_{2,p}-T_{2,S}=\mbox{FWHM}/(2\sqrt{\ln 2})$. 
}
\label{FWHM_dep_short3_HCN}
\end{figure}

The FFF, whose peak amplitude is proportional to 1/FWHM, can degrade the efficiency of the control when the FWHM of the laser pulses is too short.
And an increase of the field strengths in a simple STIRAP process does not generate greater population transfer efficiency because those stronger fields also 
generate greater interference between the active subset of states and the background states \cite{Masuda-Rice2}.
It is seen in Fig. \ref{FWHM_dep_short3_HCN} that the FFF with $\lambda=0.3$ accompanied with the STIRAP fields generates higher fidelity than
does the ordinary STIRAP and the STIRAP+FFFs with $\lambda=1$  for 8.5 ps $\le$ FWHM $\le$ 34 ps.
The drop of fidelity when FWHM decreases is due to the large intensity of the FFF.
The fields associated with STIRAP and FFF generated population transfer are complementary if the amplitude of the FFF
is not too large. 

\section{Vibrational energy transfer in thiophosgene}
The calculations reported in Section \ref{HCN} show that the efficiency of STIRAP+FFF generated population transfer when 
8.5 ps $\le$ FWHM $\le$ 34 ps is comparable to or less than that of STIRAP+CDF generated population transfer for the same pulse width range despite the
extra flexibility of STIRAP+FFF compared to STIRAP + CDF contributed by phase tuning in the former.
In this Section we show that that extra flexibility of STIRAP+FFF indeed can generate more efficient population transfer than STIRAP+CDF in the small FWHM regime, using as an example state-to-state vibrational energy transfer in nonrotating SCCl$_2$.

\begin{figure*}[]
\begin{center}
\includegraphics[width=17.5cm]{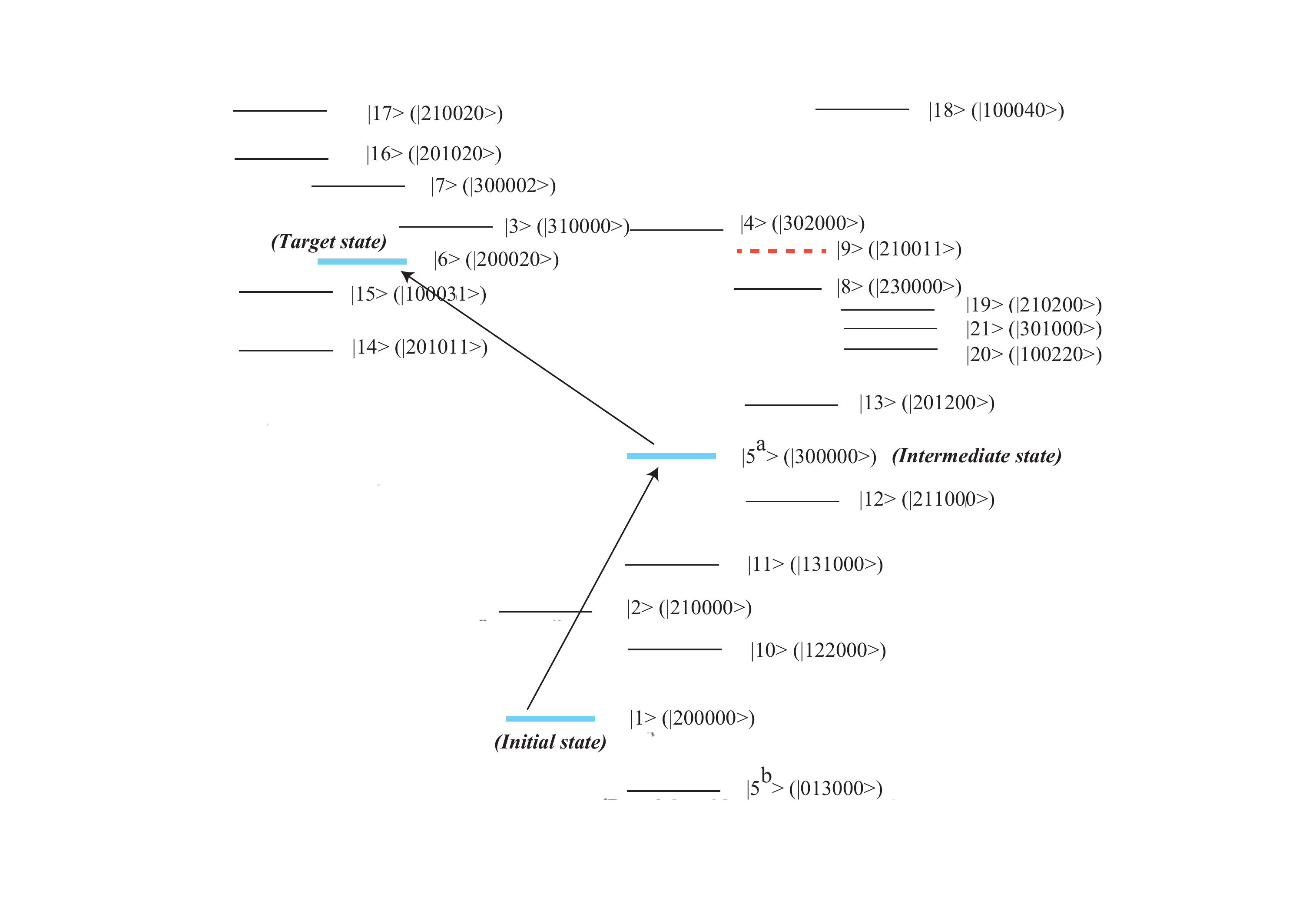}
\end{center}
\caption{
Schematic diagram of the vibrational spectrum of SCCl$_2$.}
\label{levels_fig}
\end{figure*}

The SCCl$_2$ molecule has three stretching ($\nu_1, \nu_2, \nu_3$) and three bending ($\nu_4, \nu_5, \nu_6$) vibrational degrees of freedom; 
it suffices, for our purposes, to use the same set of energies and transition dipole moments as used by Kurkal and Rice\cite{Kurkal1}, covering the
range 0 $-$ 21,000 cm$^{-1}$, determined by Bigwood, Milam and Gruebele\cite{Bigw}. These energy levels are displayed in Fig. \ref{levels_fig} and tabulated in Ref. 11.
We will focus attention on the efficiency with which population transfer can be selectively directed to one of a pair of nearly degenerate states in the presence of background states.
As in Ref. 24, we consider a STIRAP process within the subset of three states ($|200000\rangle, |300000\rangle, |200020\rangle$) embedded in the full manifold of states.  Hereafter we refer to these three states as  $|1\rangle$, $|5^a\rangle$ and $|6\rangle$, respectively.  The STIRAP+FFF control process is intended to generate higher population transfer from $|200000\rangle$ to $|200020\rangle$.  We note that $|210011\rangle$, hereafter called $|9\rangle$, with energy 5658.1828 cm$^{-1}$, is nearly degenerate with $|6\rangle$, with energy 5651.5617 cm$^{-1}$, and that the transition moment coupling states $|1\rangle$ and $|6\rangle$ is one order of magnitude smaller than those coupling states $|1\rangle$ and $|9\rangle$ and $|5^a\rangle$ and $|9\rangle$.

\begin{figure}[h!]
\begin{center}
\includegraphics[width=8cm]{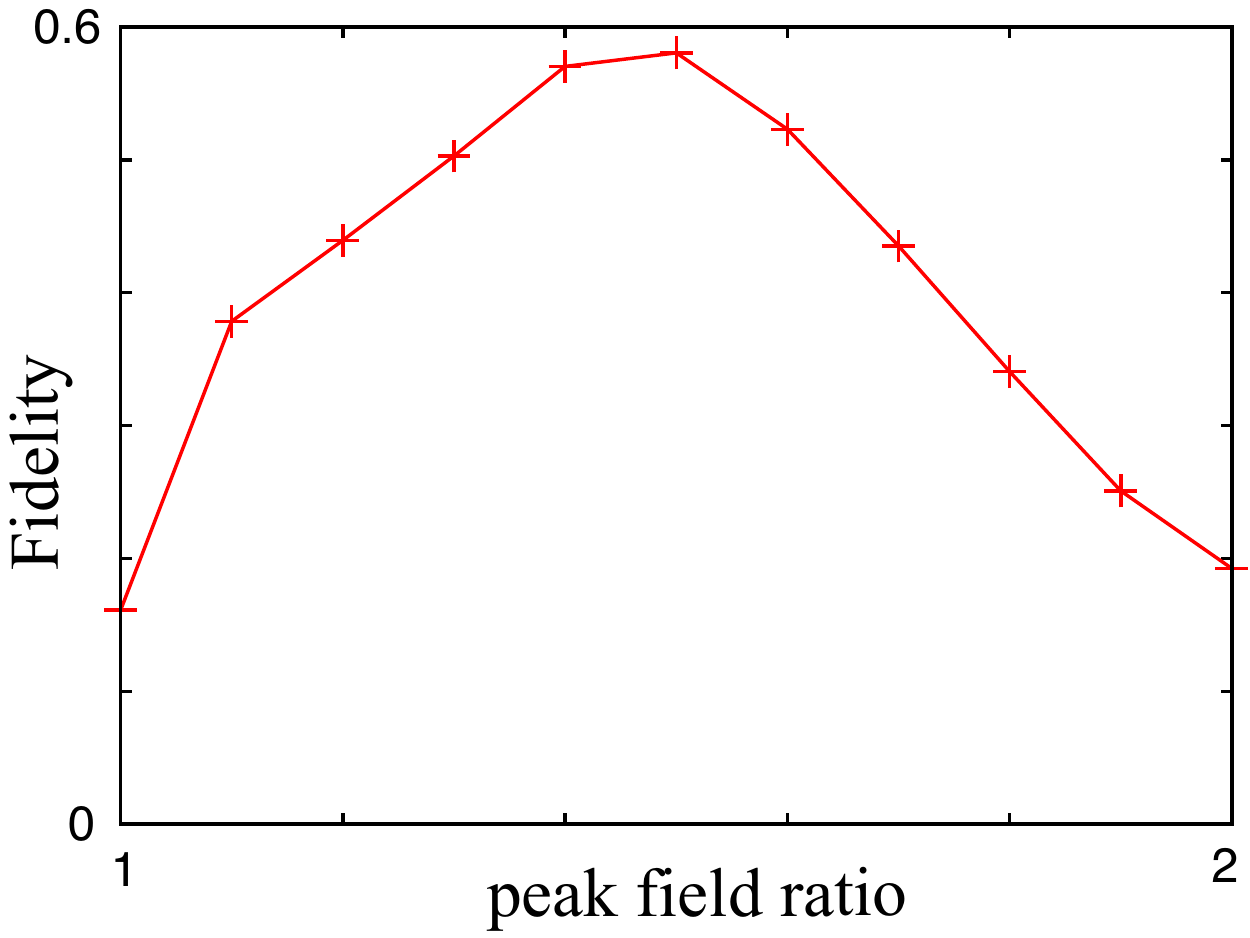}
\end{center}
\caption{
The dependence of the fidelity on the peak field ratio of the FFF to the CDF for the STIRAP+FFF control with FWHM$=21.5$ ps, $T_{p}-T_{S}=\mbox{FWHM}/(2\sqrt{\ln 2})$, 
$\lam=1$, $\tilde{E}_{p}=0.014872$ a.u. and $\tilde{E}_{S}=0.046$ a.u.
}
\label{fid_ratio}
\end{figure}

\begin{figure}[h!]
\begin{center}
\includegraphics[width=8cm]{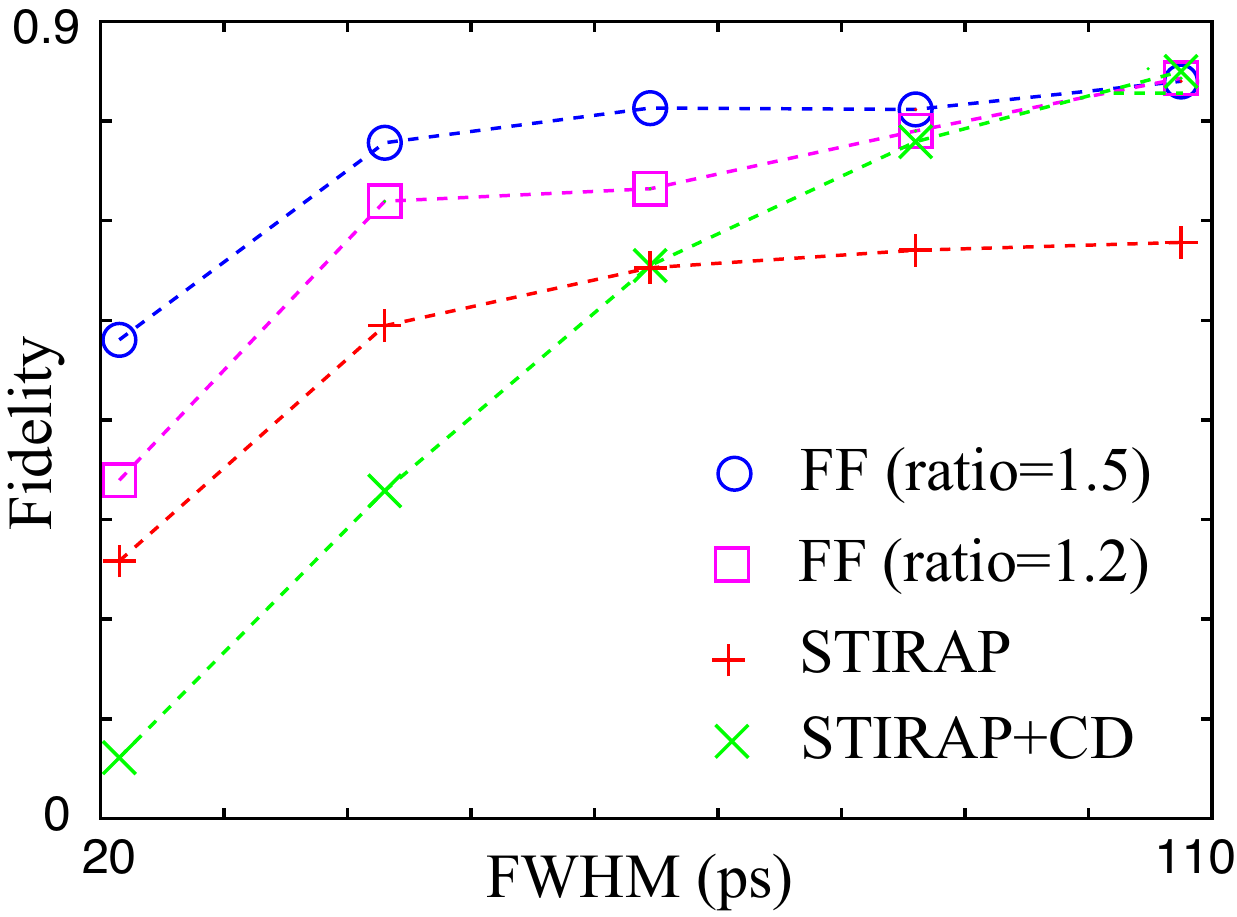}
\end{center}
\caption{
The dependence of the fidelity on the FWHM for the ordinary STIRAP, the STIRAP+CDF control and
the STIRAP+FFF with $\lambda=1$, $T_{p}-T_{S}=\mbox{FWHM}/(2\sqrt{\ln 2})$ and the peak field ratio $=1.2$ and 1.5.
}
\label{fid_FWHM_SCCl2}
\end{figure}

To compare the efficiency of the STIRAP+FFF control to that of the STIRAP+CDF control we examine 
the dependence of the STIRAP+FFF generated population transfer on the peak ratio of the FFF to the CDF.
The range of the phase tuned in the STIRAP+FFF control increases when the peak field ratio becomes large, while the population transfer when the peak field ratio is one is identical to that generated by STIRAP+CDF without phase tuning. 
Figure \ref{fid_ratio} displays the dependence of the STIRAP+FFF generated population transfer on the peak ratio of the FFF to the CDF for the parameters
FWHM$=21.5$ ps, $T_{p}-T_{S}=\mbox{FWHM}/(2\sqrt{\ln 2})$, 
$\lam=1$, $\tilde{E}_{p}=0.014872$ a.u. and $\tilde{E}_{S}=0.046$ a.u.
For a wide range of peak field ratio the population transfer generated by STIRAP+FFF exceeds that generated by STIRAP+CDF.
Figure \ref{fid_FWHM_SCCl2} displays the dependence of population transfer generated by STIRAP,
STIRAP+CDF and STIRAP+FFF (with parameters $\lam=1$ and peak field ratios 1.2 and 1.5) on the FWHM of the pulses.
The values of $\tilde{E}_{p,S}$ for each value of the FWHM have been adjusted so that the pulse areas of the pump and Stokes fields are the same as used for the 
calculations shown in Fig. \ref{fid_ratio}.
The STIRAP+FFF generated population transfer exceeds those generated by ordinary STIRAP and STIRAP+CDF for 21.5 ps $\le$ FWHM $\le$ 86 ps. 

\section{Concluding remarks}
We have examined the efficiency of STIRAP + FFF generated selective state-to-state population transfer in the vibrational manifolds of nonrotating SCCl$_2$ and the HCN$\rightarrow$CNH isomerization.  Neglecting the influence of molecular rotation on the efficiency of vibrational population transfer defines useful models that permit qualitative investigation of the influence of background states on the efficiency of energy transfer within an embedded subset of states, but those models are inadequate for the quantitative description of energy transfer in the corresponding real molecules.  It is relevant to ask if our calculations provide a qualitatively valid picture applicable to real situations.

We have argued elsewhere \cite{Masuda-Rice2} that, neglecting higher order effects such as vibration rotation interaction, we expect the rotation of a molecule to affect the state-to-state process we describe in two ways.  First, the transition dipole moment projection along the field axis differs with rotational state, thereby reducing the rate of excitation.  Second, the rotational wave-packet created may dephase on a time scale that is comparable with the width of the exciting field, thereby changing the dynamics of the population transfer.  If the ratio of the driving field duration to the period of molecular rotation is very small we expect molecular rotation to have negligible influence on the population transfer, and when the period of molecular rotation is comparable to the width of the field pulses that drive the population transfer we must expect less efficient transfer than predicted for the non-rotating molecule.  Indeed, noting that the combined STIRAP + FFF control process we describe involves both one and two photon transitions, and that the wave-packets of rotational states created by these two excitation processes have different dephasing rates, we expect the evolution of the state of the excited molecule to be complicated when the period of rotation and the exciting field duration are comparable.

The rotational periods of SCCl$_2$ and HCN are of the order of 200 ps and 10 ps, respectively.  Our calculations of the efficiency of state-to-state population transfer in SCCl$_2$ include cases when the FWHM of the pulsed fields is considerably smaller than 200 ps (see Fig. \ref{fid_FWHM_SCCl2}).  The efficiency of the population transfer is smaller when the FWHM of the pulses is 20 ps than when it is 100 ps, but still usefully large.  And since these pulse widths are of order one tenth of the rotational period it is plausible that similar efficiency of state-to-state population transfer can be achieved in the real molecule.  Our calculations of the efficiency of state-to-state population transfer in the HCN$\rightarrow$CNH isomerization do not include cases when the FWHM of the pulsed fields is considerably smaller than the rotational period.  In this case, as shown in Fig. \ref{FWHM_dep_short3_HCN}, the use of very short pulses severely degrades the population transfer efficiency.

Returning to the model cases considered, we have shown that STIRAP + FFF generated state-to-state population transfer is more efficient than STIRAP generated state-to-state population transfer when applied to a subset of states embedded in and coupled to a larger manifold of states.  Moreover, we have shown 
that the FFF calculated for an isolated subset of three states can be used to approximate the FFF applicable to a three state subset embedded in a large manifold of states even when some of the background states are strongly coupled to the intermediate state of the STIRAP process.

The FFF is designed to avert unwanted non-adiabatic population transfer at the end of the application of the pulsed field, and it directly couples the initial state to the target state thereby decreasing the sensitivity of the population transfer to the influence of background states.  STIRAP + CDF generated population transfer exhibits this same decreased sensitivity for the same reason.  However, the pulse area of the FFF is larger than $\pi$, in contrast to the pulse area of the CDF for a STIRAP + CDF process in the same system, which is always $\pi$.  And the STIRAP + FFF generated population transfer has, relative to STIRAP + CDF population transfer, an extra control parameter, namely the FFF amplitude.  This parameter can be tuned to optimize the yield of population in a target state.  In general, our model calculations show that, when the driven system of states is embedded in a large manifold of states, phase controlled STIRAP + FFF generates more efficient state-to-state population transfer than does STIRAP.

\begin{acknowledgments}
S.M. thanks the Grants-in-Aid for Centric Research of 
Japan Society for Promotion of Science and the JSPS Postdoctoral 
Fellowships for Research Abroad for its financial support.
\end{acknowledgments}



\begin{thebibliography}{}
\bibitem{Rice} S. A. Rice and M. Zhao, {\it Optical Control of Molecular 
Dynamics} (Wiley-Interscience, New York, 2000).


\bibitem{Shapiro} M. Shapiro and P. Brumer, {\it Principles of the Quantum 
Control of Molecular Processes} (Wiley-Interscience, New York, 2003).

\bibitem{Gaubatz} U. Gaubatz, P. Rudecki, S. Schiemann and 
K. Bergmann, J. Chem. Phys. {\bf 92}, 5363 (1990).

\bibitem{Coulston} G. W. Coulston, K. Bergmann, J. Chem. Phys. 
{\bf 96}, 3467 (1992).

\bibitem{Halfmann} T. Halfmann and K. Bergmann, J. Chem. Phys. 
{\bf 104}, 7068 (1996).

\bibitem{Bergmann} K. Bergmann, H. Theuer and B. W. Shore, Rev. 
Mod. Phys. {\bf 70}, 1003 (1998).

\bibitem{Vitanov} N. V. Vitanov, T. Halfmann, B. W. Shore and 
K. Bergmann, Annu. Rev. Phys. Chem. {\bf 52}, 763 (2001). 

\bibitem{Kobrak1} M. N. Kobrak and S. A. Rice, Phys. Rev. A {\bf 57}, 
1158 (1998).

\bibitem{Kobrak2} M. N. Kobrak and S. A. Rice, Phys. Rev. A {\bf 57}, 
2885 (1998).

\bibitem{Kobrak3} M. N. Kobrak and S. A. Rice, J. Chem. Phys. {\bf 109}, 
1 (1998). 

\bibitem{Kurkal1} V. Kurkal and S. A. Rice, J. Phys. Chem. B, {\bf 105}, 
6488 (2001).

\bibitem{Kurkal2} V. Kurkal and S. A. Rice, Chem. Phys. Lett. {\bf 344}, 
125 (2001).

\bibitem{Torosov} B. T. Torosov and N. V. Vitanov, Phys. Rev. A {\bf 87}, 
043418 (2013).

\bibitem{Demirplak1} M. Demirplak and S. A. Rice, J. Phys. Chem. A {\bf 107}, 
9937 (2003).

\bibitem{Demirplak2} M. Demirplak and S. A. Rice, J. Phys. Chem. B {\bf 109}, 
6838 (2005).

\bibitem{Demirplak3} M. Demirplak and S. A. Rice, J. Chem. Phys. {\bf 129}, 
154111 (2008).

\bibitem{Chen} X. Chen, I. Lizuain, A. Ruschhaupt, D. Gu\'ery-Odelin and 
J. G. Muga, Phys. Rev. Lett. {\bf 105}, 123003 (2010).

\bibitem{Muga} J. G. Muga, X. Chen, A. Ruschhaupt, and D. Gu\'ery-Odelin, 
J. Phys. B {\bf 42}, 241001 (2009).

\bibitem{Masuda1} S. Masuda and K. Nakamura, Phys. Rev. A {\bf 78}, 
062108 (2008).

\bibitem{Masuda2} S. Masuda and K. Nakamura, Proc. R. Soc. A {\bf 466}, 
1135 (2010).

\bibitem{Masuda3} S. Masuda and K. Nakamura, Phys. Rev. A {\bf 84}, 
043434 (2011).

\bibitem{Masuda4} S. Masuda, Phys. Rev. A {\bf 86}, 063624 (2012). 

\bibitem{Masuda-Rice1} S. Masuda and S. A. Rice, Phys. Rev. A {\bf 89}, 
033621 (2014). 

\bibitem{Masuda-Rice2} S. Masuda and S. A. Rice, arXiv:1409.0784 (2014).

\bibitem{Takahashi} K. Takahashi, Phys. Rev. A, {\bf 89}, 042113 (2014).

\bibitem{Cheng} T. Cheng, H. Darmawan and A. Brown, Phys. Rev. A {\bf 75}, 
013411 (2007).

\bibitem{Jakubetz} W. Jakubetz, J. Chem. Phys. {\bf 137}, 224312 (2012).

\bibitem{Smith} A. M. Smith, S. L. Coy, W. Klemperer and K. K. 
Lehmann, J. Mol. Spectrosc. {\bf 134}, 134 (1989).

\bibitem{Yang} X. Yang, C. A. Rogaski and A. M. Wodtke, J. Opt. Soc. Am. B 
{\bf 7}, 1835 (1990).

\bibitem{Jonas} D. M. Jonas, X. Yang and A. M. Wodtke, J. Chem. 
Phys. {\bf 97}, 2284 (1992).

\bibitem{Bowman} J. M. Bowman, B. Gazdy, J. A. Bentley, T. J. Lee 
and C. E. Dateo, J. Chem. Phys. {\bf 99}, 308 (1993).

\bibitem{Jakubetz2} W. Jakubetz and B. L. Lan, Chem. Phys. 
{\bf 217}, 375 (1997).

\bibitem{Bigw}
R. Bigwood, B. Milam and M. Gruebele, 
Chem. Phys. Lett. {\bf 287}, 333 (1998).


\end{thebibliography}

\end{document}